\documentclass[twocolumn,showpacs,preprintnumbers,amsmath,amssymb]{revtex4}
\usepackage{graphicx}% Include figure files
\usepackage{dcolumn}% Align table columns on decimal point
\usepackage{bm}% bold math

\begin{document}

\preprint{Submission to Phys. Rev. B}

\title{
Magnetism of the antiferromagnetic spin-$\frac{3}{2}$ dimer compound 
CrVMoO$_7$ having 
an antiferromagnetically ordered state 
}

\author{Masashi Hase$^1$}
 \email{HASE.Masashi@nims.go.jp}
\author{Yuta Ebukuro$^2$} 
\author{Haruhiko Kuroe$^2$}
\author{Masashige Matsumoto$^3$}
\author{Akira Matsuo$^4$}
\author{Koichi Kindo$^4$}
\author{James R. Hester$^5$}
\author{Taku J. Sato$^6$}
\author{Hiroki Yamazaki$^7$}

\affiliation{%
${}^{1}$Research Center for Advanced Measurement and Characterization, 
National Institute for Materials Science (NIMS), 
1-2-1 Sengen, Tsukuba-shi, Ibaraki 305-0047, Japan \\
${}^{2}$Department of Physics, Sophia University, 
7-1 Kioi-cho, Chiyoda-ku, Tokyo 102-8554, Japan \\
${}^{3}$Department of Physics, Shizuoka University, 
836 Ohya, Suruga-ku, Shizuoka-shi, Shizuoka 422-8529, Japan \\
${}^{4}$The Institute for Solid State Physics (ISSP), 
The University of Tokyo, 
5-1-5 Kashiwanoha, Kashiwa-shi, Chiba 277-8581, Japan \\
${}^{5}$Australian Centre for Neutron Scattering,  
Australian Nuclear Science and Technology Organisation (ANSTO), 
Locked Bag 2001, Kirrawee DC NSW 2232, Australia\\
${}^{6}$Institute of Multidisciplinary Research for Advanced Materials (IMRAM),
Tohoku University, 
2-1-1 Katahira, Aoba-ku, Sendai-shi, Miyagi 980-8577, Japan\\
${}^{7}$Nishina Center for Accelerator-Based Science, RIKEN, 
2-1 Hirosawa, Wako-shi, Saitama 351-0198, Japan 
}%

\date{\today}% It is always \today, today,

\begin{abstract}

We measured 
magnetization, specific heat, electron spin resonance, 
neutron diffraction, and inelastic neutron scattering 
of CrVMoO$_7$ powder. 
An antiferromagnetically ordered state 
appears below $T_{\rm N} = 26.5 \pm 0.8$ K. 
We consider that 
the probable spin model for CrVMoO$_7$
is an interacting antiferromagnetic spin-$\frac{3}{2}$ dimer model.  
We evaluated the intradimer interaction $J$
to be $25 \pm 1$ K and 
the effective interdimer interaction $J_{\rm eff}$
to be $8.8 \pm 1$ K. 
CrVMoO$_7$ is a rare spin dimer compound 
that shows an antiferromagnetically ordered state 
at atmospheric pressure and zero magnetic field.
The magnitude of ordered moments is $0.73(2) \mu_{\rm B}$.  
It is much smaller than a classical value $\sim 3 \mu_{\rm B}$. 
Longitudinal-mode magnetic excitations 
may be observable in single crystalline CrVMoO$_7$.

\end{abstract}

%\pacs{75.25.-j, 75.30.Cr, 75.40.Cx, 75.10.Jm}
%\pacs{75.25.-j, 75.30.Cr, 75.40.Cx, 75.10.Jm, 75.30.Ds}

\maketitle

\section{Introduction}

% 1st paragraph

Two types of magnetic excitations exist in a magnetically ordered state. 
They are 
gapless transverse-mode (Nambu-Goldstone mode) \cite{Goldstone62}
and 
gapped longitudinal-mode (Higgs mode) \cite{Sachdev11,Podolsky11} 
excitations 
corresponding to fluctuations in directions perpendicular and parallel 
to ordered moments, respectively. 
The transverse-mode (T-mode) excitations 
are well known as spin wave excitations.  
There are a few experimental observations 
on the longitudinal-mode (L-mode) excitations 
mainly because of their weak intensity. 
The L-mode excitations were observed in 
a pressure-induced or magnetic-field-induced 
magnetically ordered state of 
interacting antiferromagnetic (AF) 
spin-$\frac{1}{2}$ dimer compounds 
TlCuCl$_3$ \cite{Kuroe08,Ruegg08,Merchant14,Matsumoto04,Matsumoto08a}
and KCuCl$_3$ \cite{Kuroe12}. 
The ground state (GS) is a spin-singlet state 
at atmospheric pressure and zero magnetic field
in these compounds. 
The ordered state 
is in the vicinity of quantum phase transition. 
Therefore, the L-mode excitations are observable 
because of large quantum fluctuations.

% 2nd paragraph

According to results of theoretical investigations, 
the L-mode excitations may be observed 
in an antiferromagnetically ordered state 
appearing on cooling  
at atmospheric pressure and zero magnetic field 
in interacting AF spin-cluster compounds 
\cite{Matsumoto10}. 
A shrinkage of ordered magnetic moments by quantum fluctuations 
leads to a large intensity of the L-mode excitations. 
If the GS of the corresponding isolated spin cluster is a spin-singlet state, 
the shrinkage of ordered moments can be expected in an ordered state 
generated by the introduction of intercluster interactions.

% 3rd paragraph

In interacting spin clusters, 
the ordered state may appear under the condition that 
the value of an effective intercluster interaction 
is not so small compared with 
that of $\Delta$ \cite{Matsumoto10}.  
Here the effective intercluster interaction is given by the sum of 
the products of the absolute value of each intercluster interaction 
($|J_{{\rm int}, i}|$) and 
the corresponding number of interactions per spin ($z_i$) 
as $J_{\rm eff} = \sum_i z_i |J_{{\rm int}, i}|$. 
$\Delta$ is the energy difference (spin gap)
between the singlet GS and first-excited triplet states.  
It is advantageous for the appearance of the ordered state
that $\Delta$ is much smaller 
than the dominant intracluster interactions. 
In a spin-$\frac{1}{2}$ tetramer of which
the Hamiltonian  
${\cal H} = 
J_1 S_{2} \cdot S_{3} + J_2 (S_{1} \cdot S_{2}+ S_{3} \cdot S_{4})$ 
with $J_1 >0$ and $J_2 <0$, 
the GS is a spin-singlet state and 
$\Delta / J_1$ can be sufficiently small \cite{Hase97,Hase09,Hase16}.  
The values of $J_1$, $J_2$, $J_{\rm eff}$, and $\Delta$ are 
317, -162, 42, and 19 K, respectively, 
in Cu$_2$CdB$_2$O$_6$ \cite{Hase15} and 
240, -142, 30, and 17 K, 
respectively, in CuInVO$_5$ \cite{Hase16}. 
The ordered state appears in  
Cu$_2$CdB$_2$O$_6$ \cite{Hase09,Hase05,Hase15} 
and CuInVO$_5$ \cite{Hase16}
below the transition temperature 
$T_{\rm N} = 9.8$ and 2.7 K, respectively. 
Magnetic excitations in Cu$_2^{114}$Cd$^{11}$B$_2$O$_6$ 
were studied by inelastic neutron scattering 
experiments on its powder \cite{Hase15}. 
The results suggest the existence of the L-mode excitations. 
Magnetic excitations in CuInVO$_5$ 
have not been investigated.

% 4th paragraph

Spin dimer compounds are also attractive 
for investigation of the L-mode excitations 
at atmospheric pressure and zero magnetic field.  
In contrast with the small $\Delta / J_1$ in the spin-$\frac{1}{2}$ tetramer, 
the value of $\Delta /J$ is 1 
in the isolated AF spin dimer given by $J S_1 \cdot S_2$ 
irrespective of the spin value. 
It is rare that 
spin dimer compounds show a magnetically ordered state 
at atmospheric pressure and zero magnetic field.
An example is the AF spin-$\frac{1}{2}$ dimer compound
NH$_4$CuCl$_3$  
\cite{Kurniawan99,Matsumoto15,Leuenberger85}. 

% 5th paragraph

We can expect 
an interacting AF spin-$\frac{3}{2}$ dimer model 
in CrVMoO$_7$ from its crystal structure 
as shown in Fig. 1(a) \cite{Wang98,Knorr98}. 
Only the Cr$^{3+}$ ion ($3d^3$) has a localized spin-$\frac{3}{2}$. 
The shortest distance between two Cr$^{3+}$ ions 
is 3.01 \AA \ at 153 K, whereas 
the other Cr-Cr distances are larger than 
4.96 \AA \ \cite{Wang98}. 
We found an antiferromagnetically ordered state
below $T_{\rm N} = 26.5 \pm 0.8$ K. 
We investigated magnetism of CrVMoO$_7$ using 
magnetization, specific heat,  electron spin resonance, 
neutron diffraction, and inelastic neutron scattering experiments.
In this paper, we report the results. 

\begin{figure}
\begin{center}
\includegraphics[width=8cm]{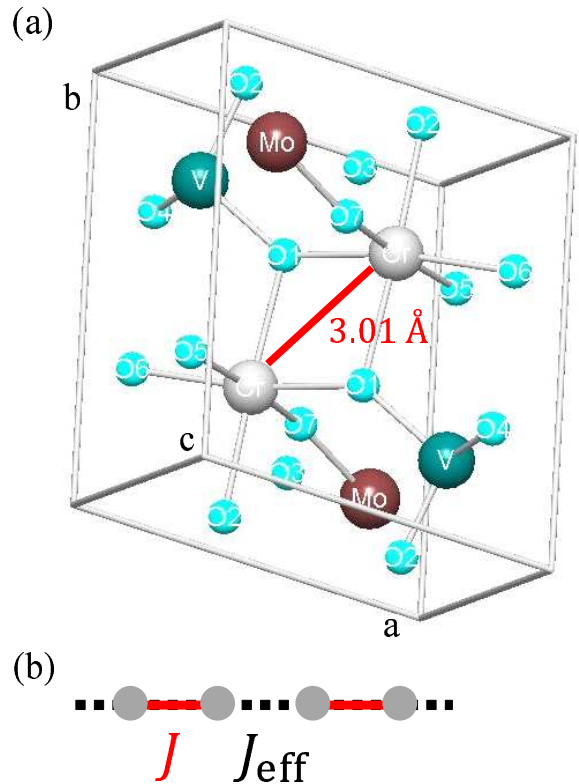}
\caption{
(Color online)
(a)
The unit cell of CrVMoO$_7$ \cite{Wang98,Knorr98}. 
An AF spin-$\frac{3}{2}$ dimer is formed by 
two neighboring Cr$^{3+}$ ions 
($3d^3$ electron configuration)
with the distance of 3.01 \AA. 
(b)
Interacting spin dimer model used to calculate magnetization 
using a mean-field theory based on the dimer unit 
(dimer mean-field theory). 
}
\end{center}
\end{figure}

\section{Experimental and Calculation Methods}

Crystalline CrVMoO$_7$ powder was synthesized 
by a solid-state reaction. 
Starting materials are 
Cr$_2$O$_3$, V$_2$O$_5$, and MoO$_3$ powder. 
Their purity is $99.99$ \%. 
A stoichiometric mixture of powder was sintered 
at 923 K in air for 268 h with intermediate grindings. 
We measured an x-ray powder diffraction pattern at room temperature 
using an x-ray diffractometer (RINT-TTR III, Rigaku). 
We confirmed that our sample was a nearly single phase of CrVMoO$_7$.  

Electron spin resonance (ESR) measurements were performed 
using an X-band spectrometer (JES-RE3X, JEOL) 
at room temperature. 
We measured the specific heat using  
a physical property measurement system (Quantum Design). 
We measured the magnetization in magnetic fields of up to 5 T 
using a superconducting quantum interference device magnetometer 
magnetic property measurement system (Quantum Design). 
High-field magnetization measurements were conducted 
using an induction method with a multilayer pulsed field magnet
installed at the Institute for Solid State Physics (ISSP), 
the University of Tokyo. 

We carried out neutron powder diffraction experiments 
using the high-intensity powder diffractometer Wombat 
(Proposal ID P5174)
at Australia's Open Pool Australian Lightwater (OPAL) reactor
in Australian Centre for Neutron Scattering 
in Australian Nuclear Science and Technology Organisation (ANSTO). 
We performed Rietveld refinements of 
the crystal and magnetic structures
using the {\tt FULLPROF SUITE} program package~\cite{Rodriguez93}  
with its internal tables for 
scattering lengths and magnetic form factors. 
We performed inelastic neutron scattering (INS) measurements 
using the inverted geometry time-of-flight spectrometer LAM-40 
in High Energy Accelerator Research Organization (KEK). 

We obtained the eigenenergies of 
isolated spin-$\frac{3}{2}$ dimers   
using an exact diagonalization method. 
We calculated 
the temperature $T$ dependence of the magnetic susceptibility $\chi (T)$
and 
the magnetic-field $H$ dependence of the magnetization $M(H)$  
using the eigenenergies. 

We calculated $M(H)$ for the model shown in Fig. 1(b) 
using a mean-field theory based on the dimer unit 
(dimer mean-field theory).
Finite magnetic moments were initially assumed on the Cr sites in the dimer.
The mean-field Hamiltonian was then expressed by a $16 \times 16$ matrix form
under consideration of the external magnetic field and 
the molecular field from the nearest neighbor sites. 
The eigenstates of the mean-field Hamiltonian were used 
to calculate the expectation value of the ordered moments on the Cr sites.
We continued this procedure until the values of the magnetic moments converged.
We finally obtained a self-consistently determined solution for $M(H)$. 

\section{Results and discussion}

% Fig. 2 EPR

Figure 2 shows the $H$ derivative of the intensity of 
electron paramagnetic resonance (EPR) 
of a CrVMoO$_7$ pellet at room temperature. 
The frequency of the incident microwave is 9.455 GHz. 
A clear resonance appeared. 
We evaluated the $g$ value to be $1.92 \pm 0.02$. 

\begin{figure}
\begin{center}
\includegraphics[width=8cm]{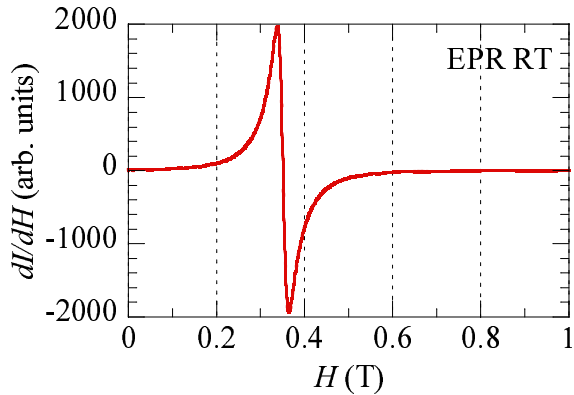}
\caption{
(Color online)
The electron paramagnetic resonance (EPR) spectrum 
of a CrVMoO$_7$ pellet at room temperature 
measured using an X-band electron spin resonance (ESR). 
}
\end{center}
\end{figure}

% Fig. 3(a) specific heat and susceptibility

Figure 3(a) shows the $T$ dependence of 
the specific heat $C(T)$ of CrVMoO$_7$ in zero magnetic field
and the $T$ derivative of the magnetic susceptibility $\chi (T)$
of CrVMoO$_7$ in $H = 0.1$~T. 
The sample was a pressed pellet and powder for 
$C(T)$ and $\chi (T)$, respectively. 
We can see a peak around 26.5 K in $C(T)$ and 
around 25.5 K in $d \chi (T)/dT$. 
As described later, 
we observed an antiferromagnetically ordered state 
at low $T$ in neutron powder diffraction experiments. 
The peak indicates the phase transition.  
We determined the transition temperature 
$T_{\rm N} = 26.5 \pm 0.8$ K  
mainly from the specific heat result. 

\begin{figure}
\begin{center}
\includegraphics[width=8cm]{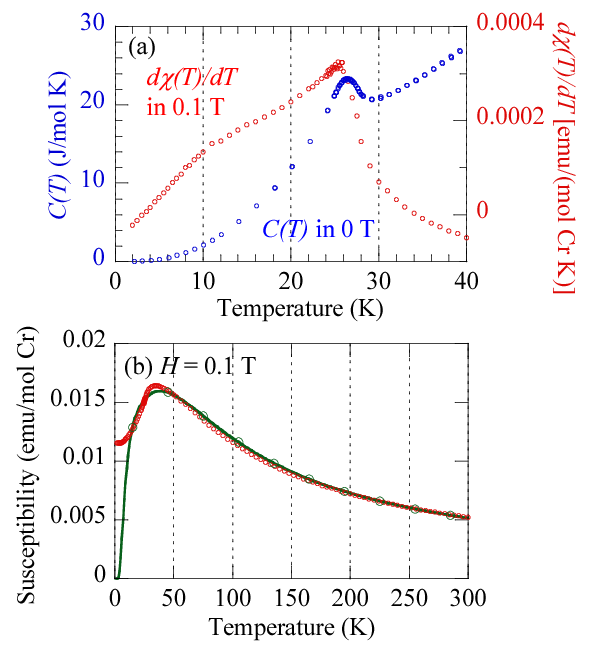}
\caption{
(Color online)
(a)
Temperature $T$ dependence of 
the specific heat $C(T)$ of CrVMoO$_7$ in zero magnetic field
and $T$ derivative of the magnetic susceptibility $\chi (T)$
of CrVMoO$_7$ 
in a magnetic field of $H = 0.1$~T. 
(b)
$T$ dependence of $\chi (T)$ of CrVMoO$_7$ 
in $H = 0.1$ \ T (red circles). 
A green line with several circles indicates 
$\chi (T)$ calculated for 
an isolated AF spin-$\frac{3}{2}$ dimer 
with $J = 25$ K and $g = 1.92$.
}
\end{center}
\end{figure}

% Fig. 3(b) susceptibility

The red circles in Fig. 3(b) show the $T$ dependence of 
$\chi (T)$ of CrVMoO$_7$ powder in $H = 0.1$~T. 
The broad maximum of $\chi (T)$ around 35 K indicates 
a low-dimensional AF spin system. 
The susceptibility seems to approach 
a finite value ($\sim 0.012$ emu/mol Cr) at 0 K. 
The magnetic order results in 
the probable finite susceptibility at 0 K. 
The susceptibility obtained by us is close to 
that reported in literature \cite{Wang98,Botto98}. 

% dimer model

We considered 
the simple isolated AF spin-$\frac{3}{2}$ dimer model 
as a first approximation 
because of the following reasons.
The spin-$\frac{3}{2}$ on Cr$^{3+}$ ions is usually 
a Heisenberg spin. 
The Cr$^{3+}$ ion is coordinated octahedrally by six oxygen ions. 
Symmetries of crystal fields affecting the Cr$^{3+}$ ions are nearly cubic. 
It is inferred that single ion anisotropy of the Cr$^{3+}$ ions is small. 
The green line in Fig. 3(b) shows $\chi (T)$ calculated for 
the isolated AF spin-$\frac{3}{2}$ dimer with $J = 25$ K and 
$g = 1.92$ that was determined in EPR. 
The experimental and calculated $\chi (T)$ 
are close to each other at high $T$. 
We evaluated $J$ to be $25 \pm 1$ K. 

An exchange interaction between two $\frac{3}{2}$ spins 
localized on magnetic ions with the $3d^3$ electron configuration
is dominated by an AF direct exchange interaction.  
Therefore, the magnitude of the exchange interaction $J_{3d^3}$
is mainly determined by the distance $R$ between two magnetic ions. 
There is an empirical relation $J_{3d^3} = a \exp (-R/b)$ 
with $a = 8.7 \times 10^{7}$ K and $b = 0.21$ \AA \
for compounds including Cr$^{3+}$ ions ($3d^3$) \cite{Hase14}. 
The value of $J_{3d^3}$ was calculated to be 53 K 
for $R = 3.01$ \AA. 
The values of $J$ and $J_{3d^3}$ are the same in order.  

% Fig. 4 M(H)

The red lines in Fig. 4 show the $H$ dependence of 
the magnetization $M(H)$ of CrVMoO$_7$ powder
measured at 1.3 and 30 K. 
The magnetization 
increases monotonically with increase in $H$. 
The green lines in Fig. 4 indicate 
$M(H)$ calculated for the isolated AF spin-$\frac{3}{2}$ dimer 
with $J = 25$ K and $g = 1.92$.  
The calculated $M(H)$ is close to the experimental $M(H)$ at 30 K, 
whereas 
the isolated spin dimer model fails to reproduce 
the experimental $M(H)$ at 1.3 K. 
There are $\frac{1}{3}$ and $\frac{2}{3}$ quantum magnetization plateaus 
in the calculated line, 
whereas 
no plateau exists in the experimental line. 
The $\frac{1}{3}$ and $\frac{2}{3}$ magnetization-plateau phases are 
polarized paramagnetic phases in which 
$S^{\rm T} = 1$ and 2, respectively. 
Here $S^{\rm T}$ 
represents the size of the total spin of the two $S = \frac{2}{3}$ spins.
We could not find the $J$ value of  the isolated spin dimer model
that reproduced 
the experimental $M(H)$ at 1.3 K.  

\begin{figure}
\begin{center}
\includegraphics[width=8cm]{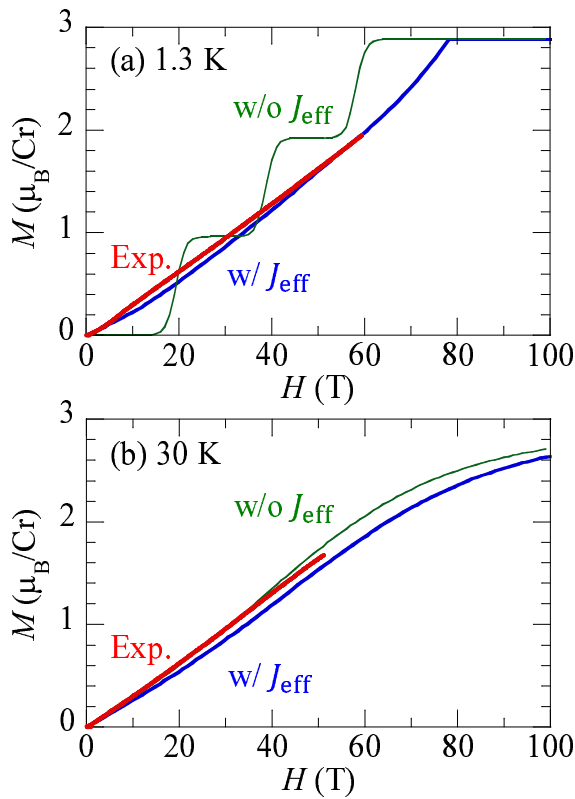}
\caption{
(Color online)
Magnetic-field $H$ dependence of 
the magnetization $M(H)$
of CrVMoO$_7$ powder (red lines). 
Blue and green lines indicate $M(H)$ calculated 
for the interacting spin-$\frac{3}{2}$ dimer model in Fig. 1(b) 
labeled by w/ $J_{\rm eff}$ and 
for an isolated spin-$\frac{3}{2}$ dimer 
labeled by w/o $J_{\rm eff}$, respectively. 
The values of the parameters are 
$J = 25$ K, $J_{\rm eff} = 8.8$ K, and $g = 1.92$. 
(a) $M(H)$ at 1.3 K. 
(b) $M(H)$ at 30 K. 
}
\end{center}
\end{figure}

According to the results in CuInVO$_5$ \cite{Hase16}, 
the discrepancy between experimental and calculated $M(H)$ 
is probably caused by interdimer interactions. 
Interdimer interactions must exist in CrVMoO$_7$
to stabilize the ordered state.
Interdimer interactions have a greater effect 
on the magnetization at lower $T$. 
Therefore, the discrepancy between the experimental results 
and those calculated for the isolated spin dimer 
appears at low $T$. 
We assumed the simple model shown in Fig. 1(b) 
as in the case of CuInVO$_5$ \cite{Hase16} 
and calculated $M(H)$ using the dimer mean-field theory. 
The blue lines in Fig. 4 indicate 
$M(H)$ calculated for the interacting spin dimer model 
with $J = 25$ K, $J_{\rm eff}=8.8$ K, and $g = 1.92$. 
The experimental and calculated $M(H)$ 
are in agreement with each other. 
The $J_{\rm eff}$ value is not so small 
compared with the $J$ value. 
Therefore, the antiferromagnetically ordered state appears. 

% Qualitative explanations of M(H)

We can explain qualitatively $M(H)$ of CrVMoO$_7$. 
In a weakly interacting spin dimer model, 
magnetization plateaus exist at low $T$. 
Magnetization-plateau phases are polarized paramagnetic phases 
without a spontaneous magnetic order.
An ordered phase can appear in a magnetic-field range, 
where $M(H)$ increases, 
between two magnetization-plateau phases. 
In the case of spin-$\frac{3}{2}$, 
there are three types of ordered phases, phase 1, 2, and 3    
at $0 \leq H < H_{1s}$, $H_{1f} < H < H_{2s}$, and 
$H_{2f} < H < H_{3s}$, respectively. 
Here, $H_{is}$ and $H_{if}$ indicate magnetic fields 
at which the $\frac{i}{3}$ plateau starts and finishes, respectively. 
The phase 1 is mainly formed by 
$S^{\rm T} = 0$ and $S^{\rm T} = 1$ states 
of isolated AF spin dimers. 
The phase 2 is mainly formed by 
$S^{\rm T} = 1$ and $S^{\rm T} = 2$ states. 
The phase 3 is mainly formed by 
$S^{\rm T} = 2$ and $S^{\rm T} = 3$ states. 
As interdimer interactions increase, 
magnetic-field ranges of ordered phases are spread. 
When interdimer interactions are strong, 
the ordered phases are connected with each other. 
A single ordered phase is formed until 
the saturation of the magnetization. 
Therefore, the magnetization increases monotonically with increase in $H$.  

% Fig. 5 crystal structure

The circles in Fig. 5 show a neutron powder diffraction pattern 
of CrVMoO$_7$ at 35 K above $T_{\rm N} = 26.5 \pm 0.8$~K. 
The wavelength $\lambda$ is 2.955 \ \AA.
We performed Rietveld refinements using 
the space group $P-1$ (No. 2) 
to evaluate crystal structure parameters. 
The line on the experimental pattern 
indicates the result of Rietveld refinements.  
The line agrees with the experimental pattern. 
The refined crystal structure parameters are presented in Table I. 
The atomic positions in our results are similar to 
those in the literature \cite{Wang98,Knorr98}.

\begin{figure}
\begin{center}
\includegraphics[width=8cm]{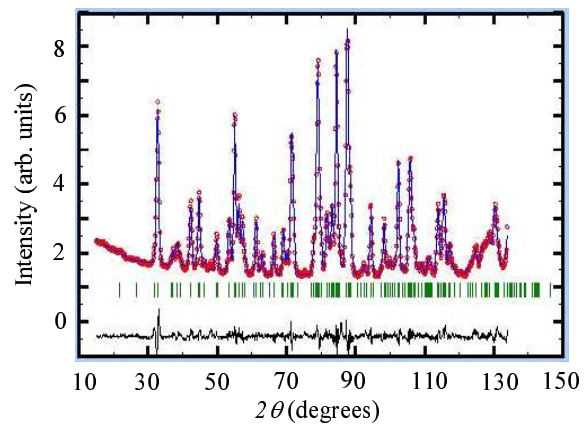}
\caption{
(Color online)
A neutron powder diffraction pattern (circles) 
of CrVMoO$_7$ at 35 K.  
The wavelength $\lambda$ is 2.955 \ \AA.  
A blue line on the measured pattern 
portrays a Rietveld refined pattern obtained using 
the crystal structure with $P-1$ (No. 2). 
A line at the bottom 
portrays the difference 
between the measured and the Rietveld refined patterns.  
Hash marks represent positions of nuclear reflections.
}
\end{center}
\end{figure}

\begin{table*}
\caption{\label{table1}
Structural parameters of CrVMoO$_7$ derived from 
Rietveld refinements of
the neutron powder diffraction pattern at 35 K. 
We used triclinic $P-1$ (No. 2).  
The lattice constants are
$a=5.521(1)$ \AA, $b=6.575(1)$ \AA, $c=7.859(1)$ \AA, 
$\alpha =96.24(1)^{\circ}$, $\beta =89.91(1)^{\circ}$, and 
$\gamma =101.99(1)^{\circ}$.  
Estimated standard deviations are shown in parentheses. 
The reliability indexes of the refinement are 
$R_{\rm p}=3.16$~\%, $R_{\rm wp}=4.14$~\%, and
$R_{\rm exp}=0.21$~\%. 
}
\begin{ruledtabular}
\begin{tabular}{cccccc}
Atom & Site & $x$ & $y$ & $z$ & $B_{\rm iso}$ \AA$^2$  \\
\hline
 Cr  & 2{\it i} & 0.826(3) & 0.310(3) & 0.408(2) & 0.30(4)\\
 V   & 2{\it i} & 0.311(3) & 0.242(3) & 0.665(3) & 0.31(5)\\
 Mo & 2{\it i} & 0.301(2) & 0.209(1) & 0.109(1) & 0.24(5)\\  
 O1 & 2{\it i} & 0.203(2) & 0.981(1) & 0.616(1) & 0.33(5)\\ 
 O2 & 2{\it i} & 0.108(3) & 0.375(1) & 0.574(1) & 0.33(5)\\ 
 O3 & 2{\it i} & 0.336(2) & 0.295(2) & 0.891(1) & 0.33(5)\\  
 O4 & 2{\it i} & 0.597(2) & 0.314(1) & 0.580(1) & 0.33(5)\\  
 O5 & 2{\it i} & 0.057(2) & 0.319(1) & 0.222(1) & 0.33(5)\\   
 O6 & 2{\it i} & 0.564(2) & 0.292(1) & 0.233(1) & 0.33(5)\\    
 O7 & 2{\it i} & 0.213(2) & 0.948(2) & 0.098(1) & 0.33(5)\\  
\end{tabular}
\end{ruledtabular}
\end{table*}

% Fig. 6&7 magnetic structure

Figure 6(a) shows neutron powder diffraction patterns 
of CrVMoO$_7$ at 5 and 35 K. 
The two patterns nearly overlap each other except for
around $2 \theta = 20^{\circ}$. 
Figure 6(b) shows the difference pattern 
made by subtracting 
the neutron powder diffraction pattern at 35 K 
from that at 5 K. 
Several magnetic reflections are apparent at 5 K.  
All the reflections can be indexed 
with the propagation vector ${\bf k} = (\frac{1}{2}, 0, \frac{1}{2})$. 

\begin{figure}
\begin{center}
\includegraphics[width=8cm]{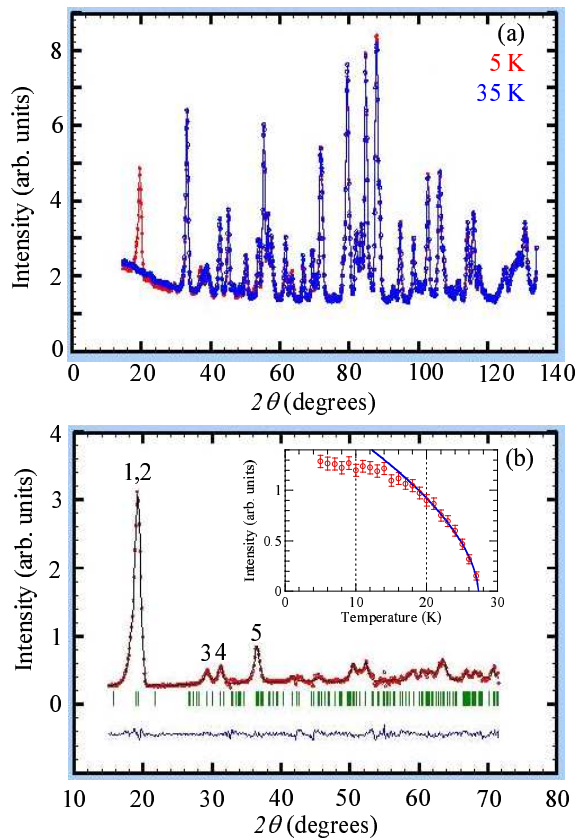}
\caption{
(Color online)
(a)
Neutron powder diffraction patterns of CrVMoO$_7$
at 5 and 35 K. 
The wavelength $\lambda$ is 2.955 \ \AA. 
(b)
A difference pattern made by subtracting 
a neutron powder diffraction pattern at 35 K 
from that at 5 K. 
A line on the measured pattern 
portrays a Rietveld refined pattern 
of the magnetic structure. 
A line at the bottom 
portrays the difference 
between the measured and the Rietveld refined patterns.  
Hash marks represent 
positions of magnetic reflections.
The reliability indexes of the refinement are 
$R_{\rm p}=4.37$~\%, $R_{\rm wp}=6.17$~\%, and
$R_{\rm exp}=0.46$~\%. 
Indexes of major magnetic reflections 
labeled by 1, 2, 3, 4, and 5
are 
$-\frac{1}{2} 0 \frac{1}{2}$,  $\frac{1}{2} 0 \frac{1}{2}$,  
$\frac{1}{2} -1 \frac{1}{2}$,  $-\frac{1}{2} 1 \frac{1}{2}$, and 
$-\frac{1}{2} 0 \frac{3}{2}$, respectively.  
The inset shows $T$ dependence of 
the integrated intensity between 17 and $22^{\circ}$.  
A blue line indicates 
$A (1- \frac{T}{T_{\rm N}})^{2 \beta}$ 
with $A = 1.98$, $T_{\rm N} = 27.3$ K, and $\beta = 0.29$. 
}
\end{center}
\end{figure}

The inset in Fig. 6(b) shows the $T$ dependence of 
the integrated intensity between 17 and $22^{\circ}$ 
including $-\frac{1}{2} 0 \frac{1}{2}$ and $\frac{1}{2} 0 \frac{1}{2}$ 
reflections. 
The intensity increases with decrease in $T$ and 
is nearly constant below 14 K. 
The blue line indicates 
$A (1- \frac{T}{T_{\rm N}})^{2 \beta}$ 
with $A = 1.98$, $T_{\rm N} = 27.3$ K, and $\beta = 0.29$. 
These values were obtained from the data above 20 K. 
We evaluated $\beta$ to be 0.26 in 
the spin-$\frac{1}{2}$ tetramer compound Cu$_2$CdB$_2$O$_6$ 
from the inset figure in Fig. 4 in \cite{Hase09}.  
The two values of the critical exponent 
are close to each other. 
The $\beta$ value is 0.36, 0.33, and 1/8 for 
three-dimensional Heisenberg, 
three-dimensional Ising, and 
two-dimensional Ising models, respectively. 
In the Ising models, 
$\beta$ is smaller in the lower dimension. 
The spin models in CrVMoO$_7$ and Cu$_2$CdB$_2$O$_6$ 
are low-dimensional AF interacting spin clusters. 
Therefore, the $\beta$ values in these compounds 
are smaller than that of three-dimensional Heisenberg models.  

According to magnetic space groups in $P-1$ \cite{Litvin08}, 
only a collinear magnetic structure is possible. 
We performed Rietveld refinements 
for the difference pattern using two models. 
Two ordered moments in each dimer are 
parallel in one model and antiparallel in the other one.  
As expected, only the antiparallel model can explain 
the magnetic reflections as shown in Fig. 6(b). 

The magnetic structure is shown in Fig. 7 \cite{Comment01}. 
An ordered moment vector is 
$(0.02(2), 0.60(1), -0.36(2)) \mu_{\rm B}$ 
lying nearly in the $bc$ plane.  
Its magnitude is $0.73(2) \mu_{\rm B}$.  
It is much smaller than a classical value $\sim 3 \mu_{\rm B}$. 
The GS of the spin dimer is a spin-singlet 
state \cite{Hase93a,Hase93b,Hase93c}. 
Therefore, the ordered moment is shrunk. 

\begin{figure}
\begin{center}
\includegraphics[width=8cm]{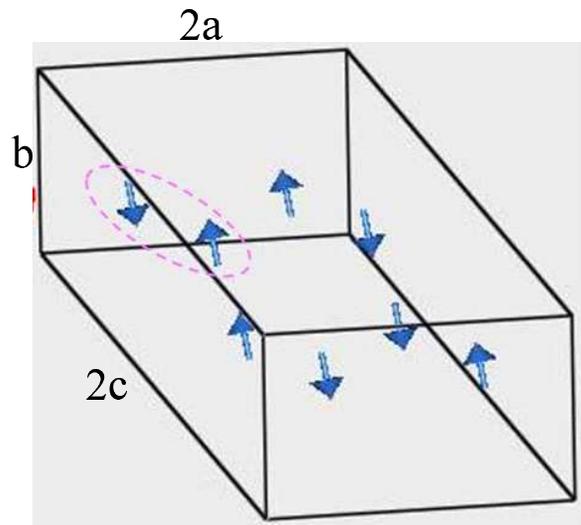}
\caption{
(Color online)
The magnetic structure of CrVMoO$_7$. 
An ellipse indicates an AF dimer.
}
\end{center}
\end{figure}

% Fig. 8 & 9 INS results

Figure 8 shows INS
intensity $I(Q, \omega)$ maps of CrVMoO$_7$ powder 
at 1.5 and 30 K. 
Here, $Q$ and $\omega$ are 
the magnitude of the scattering vector and the energy transfer, respectively. 
The energy of final neutrons $E_{\rm f}$ is 4.59 meV. 
We can see excitations between 2 and 7 meV at 1.5 K. 
The intensity of the excitations  
is suppressed at higher $Q$. 
Excitations at 1.5 K also exist below 2 meV 
around $Q = 0.7$ \AA$^{-1}$. %$
Excitations at 30 K exist in lower energies
in comparison with those at 1.5 K. 
The intensity is strong at low $\omega$ 
around $Q = 0.7$ \AA$^{-1}$. %$

\begin{figure}
\begin{center}
\includegraphics[width=8cm]{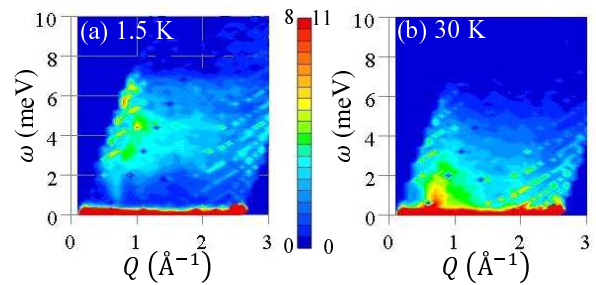}
\caption{
(Color online)
INS intensity $I(Q, \omega)$ maps 
in the $Q - \omega$ plane for 
CrVMoO$_7$ powder at 1.5 K (a) and 30 K (b) 
measured using the LAM-40 spectrometer.
The energy of final neutrons $E_{\rm f}$ is 4.59 meV. 
The right vertical key shows the INS intensity in arbitrary units. 
}
\end{center}
\end{figure}

Figure 9(a) shows 
the $\omega$ dependence of $I(Q, \omega)$ 
in the $Q$ range of $0.95 - 1.05$ \AA$^{-1}$. %$
The intensity at 1.5 K 
is the strongest around 4.5 meV. 
The intensity at 30 K 
decreases with increase in $\omega$. 
The red circles in Fig. 9(b) show 
the $Q$ dependence of $I(Q, \omega)$ at 1.5 K 
summed in the $\omega$ range of 4 - 5 meV.
The intensity shows a peak around 
$Q = 1.0$ \AA$^{-1}$. %$

\begin{figure}
\begin{center}
\includegraphics[width=8cm]{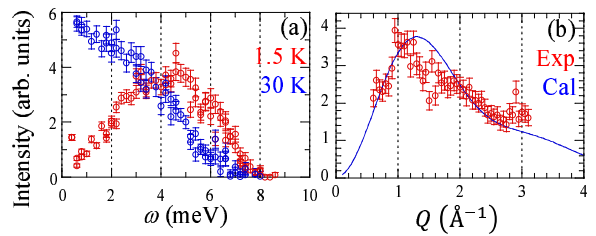}
\caption{
(Color online)
(a)
$\omega$ dependence of the INS intensity 
for CrVMoO$_7$ powder
in the $Q$ range of $0.95 - 1.05$ \AA$^{-1}$ %$
at 1.5 and 30 K.
(b)
The $Q$ dependence of the INS intensity 
of CrVMoO$_7$ powder
summed in the $\omega$ range of 4 - 5 meV 
at 1.5 K (circles).
A line indicates 
the intensity calculated for 
an isolated AF spin-$\frac{3}{2}$ dimer.  
The formula is 
$A f(Q)^2 [1- \sin (3.01 Q)/ (3.01 Q)]$ where 
$A$ and $f(Q)$ represent 
a scaling factor and an atomic magnetic form factor, 
respectively.
}
\end{center}
\end{figure}

Considering the INS results of Cu$_2$CdB$_2$O$_6$ \cite{Hase15}, 
we can explain qualitatively the INS results of CrVMoO$_7$ using 
the interacting AF spin-$\frac{3}{2}$ dimer model.  
The blue line in Fig. 9(b) indicates 
the $Q$ dependence of the intensity calculated for 
the isolated spin dimer model 
with the Cr-Cr distance of 3.01 \AA. 
The experimental and calculated results 
are similar to each other. 
The first excited spin-triplet states exist at 
2.2 meV (= 25 K)
in the isolated AF spin-$\frac{3}{2}$ dimer.  
Interdimer interactions change
discrete levels of excited states to 
excitation bands with finite widths. 
The excitations between 2 and 7 meV 
indicate the existence of  the excitation bands \cite{Comment02}.  

% gapless point

The magnetic reflection 
is the strongest at $-\frac{1}{2} 0 \frac{1}{2}$. 
The magnetic zone center of the spin configuration shown in Fig. 7 
is $-\frac{1}{2} 0 \frac{1}{2}$.
The $Q$ value is 0.70 \AA$^{-1}$. %$ 
Therefore, the excitations at 1.5 K below 2 meV 
around $Q = 0.7$ \AA$^{-1}$ %$ 
are T-mode (Nambu-Goldstone mode) excitations 
in the vicinity of the gapless point.  

% 30 K

The magnetic excitations are gapless below $T_{\rm N}$. 
The temperature 30 K 
is slightly higher than $T_{\rm N} = 26.5 \pm 0.8$ K. 
The bandwidths are large and  
the excitation gap is small.  
Therefore, magnetic excitations appear in low energies. 
Excitations from thermally excited states in the excitation bands 
also generate the continuous low-energy intensities at 30 K.

% last paragraph

We could not confirm L-mode excitations 
because of the powder sample. 
We intend to make single crystals of CrVMoO$_7$ and 
to perform INS and Raman scattering experiments on them 
to investigate L-mode excitations. 
We expect that 
L-mode excitations are observable 
because of the small ordered moment. 

\section{Conclusion}

We investigated magnetism of CrVMoO$_7$ using 
magnetization, specific heat, electron spin resonance, 
neutron diffraction, and inelastic neutron scattering 
experiments. 
An antiferromagnetically ordered state 
appears below $T_{\rm N} = 26.5 \pm 0.8$ K. 
The magnetic susceptibility of CrVMoO$_7$ powder at high $T$
is close to that calculated for 
the isolated AF spin-$\frac{3}{2}$ dimer with 
the intradimer interaction value $J = 25 \pm 1$ K and 
$g = 1.92 \pm 0.02$. 
We were able to explain the magnetization curves 
using the interacting AF spin-$\frac{3}{2}$ dimer model with
the effective interdimer interaction $J_{\rm eff} = 8.8 \pm 1$ \ K.
We determined the magnetic structure of CrVMoO$_7$. 
The magnitude of ordered moments is $0.73(2) \mu_{\rm B}$.  
It is much smaller than a classical value $\sim 3  \mu_{\rm B}$. 
Two ordered moments are antiparallel
in each dimer.  
We observed magnetic excitations 
in inelastic neutron scattering experiments. 
We can explain qualitatively the results using 
the interacting AF spin-$\frac{3}{2}$ dimer model.  
CrVMoO$_7$ is a rare spin dimer compound 
that shows an antiferromagnetically ordered state 
at atmospheric pressure and zero magnetic field.
Longitudinal-mode magnetic excitations 
may be observable in single crystalline CrVMoO$_7$.

\begin{acknowledgments}

This work was financially supported by 
Japan Society for the Promotion of Science (JSPS) 
KAKENHI (Grant No. 15K05150) and 
grants from National Institute for Materials Science (NIMS). 
M. M. was supported by 
JSPS KAKENHI (Grant No. 26400332). 
The high-field magnetization experiments were conducted 
under the Visiting Researcher's Program of 
the Institute for Solid State Physics (ISSP), the University of Tokyo. 
The neutron powder diffraction experiments were performed 
by using the Wombat diffractometer at 
Australian Nuclear Science and Technology Organisation 
(ANSTO), Australia 
(proposal ID. P5174).
We are grateful to S. Matsumoto 
for sample syntheses and x-ray diffraction measurements.   

\end{acknowledgments}

\newpage %Just because of unusual number of tables stacked at end

\begin{references}
\bibitem{Comment01}
Strictly speaking, 
the unit cell vectors of the magnetic structure are 
$\vec{a}+\vec{c}$, $\vec{b}$, and $\vec{a}-\vec{c}$. 
The unit cell of the magnetic structure is 
the double of that of the crystal structure.
\bibitem{Comment02}
In the spin-$\frac{1}{2}$ tetramer compound 
Cu$_2$CdB$_2$O$_6$ \cite{Hase15},  
the first excited spin-triplet states exist at 1.6 meV 
in the isolated tetramer. 
Magnetic excitations are spread 
between 1.6 and 3.0 meV 
by intertetramer interactions.
\end{references}

\end{document}